\documentclass[twocollumn,aps, pre, reprint]{revtex4-1}

\usepackage{natbib}
\usepackage{amssymb,amsmath, amsfonts}
\usepackage{graphicx}
\usepackage{hyperref}
\usepackage[mathlines]{lineno}
\linenumbers

\usepackage[usenames,dvipsnames]{color}
\usepackage[utf8]{inputenc}
\usepackage{ulem}

\begin{document}

\preprint{DNLSC-002/2010 -- v1.2}

\title{Parametric evolution of unstable dimension variability in coupled piecewise-linear chaotic maps}

\author{R. F. \surname{Pereira}$^1$}
\author{M. C. \surname{Verg\`es}$^2$}
\author{R. L. \surname{Viana}$^2$}
\author{S. R. \surname{Lopes}$^2$}
\author{S. E. de S. \surname{Pinto}$^1$}
\email[Corresponding author: ]{desouzapinto@pq.cnpq.br}
\affiliation{1.Departamento de F\'isica, Universidade Estadual de Ponta Grossa, 84030-900, Ponta Grossa, PR, Brazil\\ 2.Departamento de F\'isica, Universidade Federal do Paran\'a, Caixa Postal 19044, 81531-990, Curitiba, Paran\'a, Brazil}
 
\date{\today}

\begin{abstract}
In presence of unstable dimension variability  numerical  solutions of chaotic systems are valid only for short periods of observation. For this reason, analytical results for systems that exhibit this phenomenon are needed. Aiming to go one step further in obtaining such results, we study the parametric evolution of unstable dimension variability in two coupled bungalow maps. Each of these maps presents intervals of linearity that define Markov partitions, which are recovered for the coupled system in the case of synchronization. Using such partitions we find exact results for the onset of unstable dimension variability and for contrast measure, which quantifies the intensity of the phenomenon  in terms of the stability of the periodic orbits embedded in the synchronization subspace.
\end{abstract}


\maketitle


Unstable dimension variability (UDV) is a  form of non-hyperbolicity in which there is no continuous splitting between stable and unstable subspaces along the chaotic invariant set \cite{PhysicaD.v109.p81.y1997}. The variability takes place when the periodic orbits, embedded in the chaotic set, have a different number of unstable directions. This  is a local phenomenon that can influence the entire phase space, and create complexity in the system \cite{Riddling.Lai.Celso,PRE.v59.pR3807.y1999,Tribolium}. Validity of trajectories generated by chaotic systems that exhibit  UDV  is guaranteed for short periods \cite{PRE.v65.p036220.y2002}, which decreases as the intensity of the UDV increases \cite{PRL.v79.p59.y1997,PintoCelsoPRE}.

The intensity of the UDV can be quantified by the embedded UPOs in a nonhyperbolic attractor \cite{PRE.v59.pR3807.y1999}. There are efficient computational methods for the analysis of these orbits \cite{PRL.v78.p4733.y1997, PRE.v60.p6172.y1999}. However, it is a time-consuming task because the number of orbits increases with their period, and in many problems it is necessary to consider very high periods \cite{PRE.v55.pR1251.y1997, PRE.v56.p4031.y1997, Chaos.v17.p023131.y2007}. To avoid this problem, one constructs a model so that the UDV occurs in a transversal direction to a hyperbolic attractor. The dynamics in this attractor is well known, and therefore, some analytical results can be obtained. This type of construction allows us develop tools in order to shed light on the UDV \cite{PLA.v270.p308.y2000, IJBC.v10.p683.y2000}. Examples of physical problems that can be handled by these tools are: the effect of shadowing in the kicked double-rotor \cite{PRL.v73.p1927.y1994, PLA.v372.p5569.y2008},  the beginning of the spatial activity in the three-waves model \cite{PhysicaD.v238.p516.y2009, SzezechJr2010}, transport properties of passive inertial particles  incompressible flows \cite{PRE.v79.p066203.y2009}, and the chaos synchronization in coupled map lattices \cite{PRL.v82.p4803.y1999, PRE.v68.p067204.y2003, PhysicaD.v206.p94.y2005}. In some cases, the study of periodic orbits embedded in the synchronization subspace allows the determination of the global behavior of coupled chaotic maps \cite{PhysicaA.v389.p5279.y2010}.

The lack of accurate results hinders the understanding of the UDV. Thus, the key question that this article will address is the analytical calculations for systems that present such phenomenon. In the following pages, we shall consider a simple spatially extended system composed by two identical bungalow maps \cite{IJTP.v37.p2653.y1998} , which are piecewise linear, and interacts by a diffusive coupling. Such a system exhibits chaos synchronization and UDV in the transversal direction to the  synchronization subspace, for certain parameters intervals \cite{PhysicaA.v388.p2515.y2009}. Besides, this map presents strong chaos for the entire parameter control interval \cite{PRE.v55.p7763.y1997}. These features allows us to study the parameter evolution of the UDV for arbitrary periods.

Now, we shall consider the abovementioned map, $x \mapsto f_a (x)$, given by 
\begin{equation}\label{EQbungalow}
	f_a(x) =\left\{
	\begin{array}{ll}
		\frac{1-a}{a}x & \mbox{if }x\in I_1 \equiv \left[0,a\right)\\
		\frac{2a}{1-2a}x+\frac{1-3a}{1-2a} & \mbox{if }x\in I_2 \equiv \left[ a,\frac{1}{2}\right)\\
		\frac{2a}{1-2a}(1-x) + \frac{1-3a}{1-2a} & \mbox{if }x\in I_3 \equiv [\frac{1}{2},1-a) \\
		\frac{1-a}{a}\left( 1-x\right) & \mbox{if }x\in I_4 \equiv \left[1-a,1\right]
	\end{array}\right.
\end{equation}
in which $a \in (0,1/2)$ is a parameter. This map has the following property \cite{IJTP.v37.p2653.y1998}: $\forall a$, the four intervals of linearity of the map define four Markov partitions \footnote{In the case where $a=1/3$, the map (\ref{EQbungalow}) is reduced to tent map, for which there are two partitions: $0\leqslant x < 1/2$ e $1/2 \leqslant x \leqslant 1$.} of phase space $\omega=\bigcup I_\alpha=[0,1]$ (going forward, greek indexes range from 1 to 4). This property allows us to study the symbolic and, consequently, the interval dynamics of the system. Therefore, we hypothesize that the Markov partions allows an exact result for the onset of the UDV in the synchronization subspace of coupled bungalow maps.

\begin{figure}[tb]
  \includegraphics[clip, width=0.4\hsize, clip]{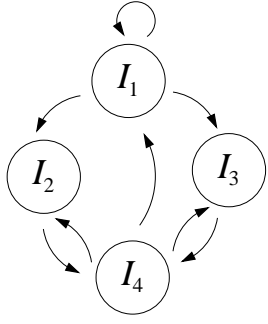}
  \caption{\label{FIGgraph} Possible transitions between partitions of the bungalow map.}
\end{figure}

In order to do this study, we must determine all possible itineraries. Considering the linearity of the map in each interval $I_\alpha$ and the images of its ends,
\begin{equation*}
	\left.
		\begin{array}{rcl}
			f_a(0) & = & 0\\
			f_a(a) & = & 1-a\\
			f_a(1/2) & = & 1\\
			f_a(1-a) & = & 1-a\\
			f_a(1) & = & 0
		\end{array}
	\right\}
	\Rightarrow
	\left\{
	\begin{array}{l}
		f_a(I_1) \subset \left(I_1 \cup I_2 \cup I_3\right)\\
		f_a(I_2) \subset I_4\\
		f_a(I_3) \subset I_4\\
		f_a(I_4) \subset \left(I_1 \cup I_2 \cup I_3\right)
	\end{array}
	\right.,
\end{equation*}
we obtain the {graph indicated} in Fig. \ref{FIGgraph}.

Consider, now, the following matrix \cite{PRE.v55.p7763.y1997, ChaosBook}:

\begin{equation}\label{EQtransfermatrix}
  T \circeq 
  \left[
  	\begin{array}{cccc}
	    \eta_1 & \eta_1 & \eta_1 & 0\\
	    0 & 0 & 0 & \eta_2\\
	    0 & 0 & 0 & \eta_3\\
	    \eta_4 & \eta_4 & \eta_4 & 0
  	\end{array}
  \right],
\end{equation}
whose eigenvalues are given by $t_{1,2} = (\eta_1\pm\theta)/2$ and $t_{3,4}=0$, in which $\theta \equiv \sqrt{\eta_1^2 + 4\eta_4(\eta_2+\eta_3)}$. 

It is straightforward to apprehend that matrix (\ref{EQtransfermatrix}), with all $\eta_\alpha=1$, represents the transfer matrix $T_1$ -- associated with the graph in Fig. \ref{FIGgraph} -- of the map, where the element located in the line $\nu$ and column $\tau$ of the $n-$th power, $[T_1^n]_{\nu\tau}$, represents the number of different itineraries of size $n$ that start in the partition $I_\nu$ and end in the partition $I_\tau$. Therefore, the topological entropy of map (\ref{EQbungalow}) is given by logarithm of the largest eigenvalue ($t_1$) of the matrix $T_1$ ($h_T = \ln 2$) \cite{ChaosBook}. Moreover, the invariant density of the map is given by the eigenvector components associated with $t_1$: $\mathbf{v}_1 = 1/\sqrt{10}[2\,1\,1\,2]^T$ (the component $v_1^{(\alpha)}$ indicates the natural measure of the $\alpha$-th partition).

In matrix (\ref{EQtransfermatrix}), the $\eta_\alpha$ stands for any quantity that is constant in each interval of linearity $I_\alpha$ and multiplicative along a trajectory. Thus, we can use the $n$-th power of the matrix (\ref{EQtransfermatrix}) to study the dynamical properties of the map. For example, the diagonal elements of $T^n$ provide the $2^n$ periodic sequences of size $n$.

The trace of the matrix is directly related to its eigenvalues by
\begin{equation*}
  \mbox{tr}T^n = \sum_\alpha t^n_\alpha.
\end{equation*}
Once we know the eigenvalues of $T$, we can determine the trace of $T^n$, whatever the value of  $n$: 

\begin{equation}\label{EQtrace}
  \mbox{tr}T^n = \frac{1}{2^n} \sum_{k = 0}^n\dbinom{n}{k}\eta_1^k\theta^{n-k}[1+(-1)^{n-k}].
\end{equation}

In Eq. (\ref{EQtrace}) each term in the summation is related to a possible symbolic sequence. Thus, if  $\eta_\alpha = f'(x)|_{x\in I_\alpha}$, then Eq. (\ref{EQtrace}) gives the stability coefficients spectrum of the $n$-th periodic points of the map (\ref{EQbungalow}). As an illustration , we have associated with the intinerary  $I_1I_2I_4I_1I_2I_4\cdots\overline{I_1I_2I_4}$ a point of period 3. For this case the coefficient of stability is the product $\eta_1\eta_2\eta_4$.

From now on we shall examine the case of two coupled maps. We shall use the following  version for the coupling: 
\begin{equation}\label{EQcoupledmaps} 
	\dbinom{x_{n+1}}{y_{n+1}}=\mathbf{G}\dbinom{x_n }{y_n }=\dbinom{f_a(x_n +\delta(y_n-x_n))}{f_a(y_n+\varepsilon(x_n-y_n))}, 
\end{equation} 
in which $\delta$ and $\varepsilon$ can take on different (asymmetric coupling) or equal values (symmetric coupling). If either of them vanishes, we obtain a master-slave coupling. In any instance, the dynamics, for synchronization purposes, will depend on their sum $d \equiv \delta+\varepsilon$. 
 
The map $\mathbf{G}$ keeps the unit square invariant when both $0 \leq \delta \leq 1$ and $0 \leq \varepsilon \leq 1$ (we will deal only with these intervals). This system have the property that the dynamics it generates leave the straight line $x=y$ of the plane invariant and, consequently, the segment $\mathcal{S}=\{(x,y)\in\omega^2\mid 0\leq x=y \leq1\}$. The latter is often called the {\it synchronization subspace}.

Since UDV is a local phenomenon, we shall consider the transversal linear stability to the synchronization subspace. So, we linearize the system (\ref{EQcoupledmaps}) and diagonalize  it-- in the basis of the Jacobian matrix -- in the directions $\mathbf{u}_\parallel = [1\;1]^T$ and $\mathbf{u}_\perp = [\delta\;-\varepsilon]^T$. The quantities associated with the directions $\mathbf{u}_\parallel$ and $\mathbf{u}_\perp$, are called longitudinal and transversal, respectively.

By definition,  $\mathcal{S}$ is nonhyperbolic if there is at least one periodic point embedded in that subspace whose unstable dimension is different from any other point in $\mathcal{S}$. By construction, all periodic points in the set are longitudinally unstable. Therefore, the phenomenon occurs in this system only in the transversal direction and, it is necessary that periodic points transversely stable and unstable  coexist with each other in $\mathcal{S}$.   In order to study, in a quantitative way, the unstable dimension variability of the system, we must determine the unstable dimension of {\it all} periodic points of the map.  We must also determine the frequency  with which a typical trajectory visits the neighborhood of these points. As in the synchronization manifold the dynamics is hyperbolic and mixing, we know that such frequency can be obtained by the invarariant density given by \footnote{ The natural measure -- generated by any typical trajectory --  of any subset $\mathbb{D}\in\omega$ is given by  $\mu(\mathbb{D})=\int_\mathbb{D}\rho(x)\mbox{d}x$.}\cite{PRA.v37.p1711.y1988}
\begin{equation}\label{EQdensity}
  \rho(x) = \frac{1}{\Delta x}\lim_{p\rightarrow\infty}\sum_{x\in\mathbb{D}}\frac{1}{|\Lambda_{\parallel}(x, p)|}, 
\end{equation}
where $\mathbb{D}= [x, x + \Delta x)$, and the summation extends over all points of period $p$ in $\mathbb{D}$, whose eigenvalues associated with the longitudinal direction\footnote{Now we consider the dynamics in the synchronization subspace.} are given by $\Lambda_{\parallel}(x, p)$. Note that expression (\ref{EQtrace}) give us all possible eigenvalues for all points fo period $p=n$. It is possible  calculate, from (\ref{EQtrace}), the number of periodic points which have the same eigenvalue. For this purpose, we rewrite $\theta^{n-k}$ as follows\footnote{Note that the term  $\left[ 1 + \left( - 1 \right)^{n - k} \right]$ in  (\ref{EQtrace}) filters only the terms $(n-k)$ which are even.}
\begin{equation*}
	\theta^{n-k} = \eta_1^{n-k} \sum_{w=0}^{\frac{n-k}{2}} \dbinom{\frac{n-k}{2}}{w} \left(\frac{2}{\eta_1}\right)^{2w}\sum_{r = 0}^w \dbinom{w}{r}\eta_2^r\eta_3^{w-r}\eta_4^w
\end{equation*}
  
Replacing in  (\ref{EQtrace}), we obtain
\begin{equation}\label{EQseries}
	\begin{array}{rrr}
		\mbox{tr}T^n & = & \displaystyle\sum_{k=0}^n \dbinom{n}{k} \left[1+(-1)^{n-k}\right] \sum_{w=0}^{\frac{n-k}{2}} \left(\frac{1}{2}\right)^{n-2w}\times\\
		& &\displaystyle \times \sum_{r=0}^w \dbinom{\frac{n-k}{2}}{w}  \dbinom{w}{r} \eta_1^{n-2w} \eta_2^r \eta_3^{w-r} \eta_4^w.
	\end{array}
\end{equation}

Equation (\ref{EQseries}) gives all information required by Eq. (\ref{EQdensity}). Since the system is piecewise-linear, the eigenvalues obtained by  $\eta_1^{n - 2 w} \eta_2^r \eta_3^{w - r} \eta_4^w$, can be used to determine in which partition $\mathbb{D}$, these periodic points are contained. Thus, taking the $\eta$'s as the trasversal eigenvalues, we determine the unstable dimension of partition  $\mathbb{D}$. On the other hand, taking the $\eta$'s as the longitudinal eigenvalues, and theirs coefficients, we determine the measure, {\it i.e.} the contribution of this partition to the behavior of typical trajectories in the vicinity of the synchronization manifold. This analysis allows us to quantify the unstable dimension variability.


First, we determine the set of parameters for which the UDV occurs. In order to calculate the beginning of this phenomenon we evaluate the coefficients of stability of each partition. Therefore, the determination of the parameters $a_c$ and $d_c=\delta+\varepsilon$, which are critical for the beginning and the end of unstable dimension variability, is done by calculating the possible transversal eigenvalues $\eta_1^{n - 2w} \eta_2^r \eta_3^{w - r} \eta_4^w$, with
\begin{eqnarray*}
  \eta_1 & = & -\eta_4 = \left(\frac{1-a}{a}\right)(1-d)\\
  \eta_2 & = & -\eta_3 = \left(\frac{2a}{1-2a}\right)(1-d).
\end{eqnarray*}

Simply we determine which possible combinations of $\eta_1^{n - 2 w}\eta_2^r\eta_3^{w-r}\eta_4^w$ result the largest $(\sup|\Lambda_{\perp}|)$ and the lowest $(\inf|\Lambda_{\perp}|)$ eigenvalues, in magnitude, and evaluate the range of existence of the UDV like follows
\begin{equation}\label{EQudvboundaries}
	\sup|\Lambda_{\perp}|=1\;\mbox{ and }\;\inf|\Lambda_{\perp}|=1.
\end{equation}

\begin{figure}[tb]
  \includegraphics[clip, width=\hsize]{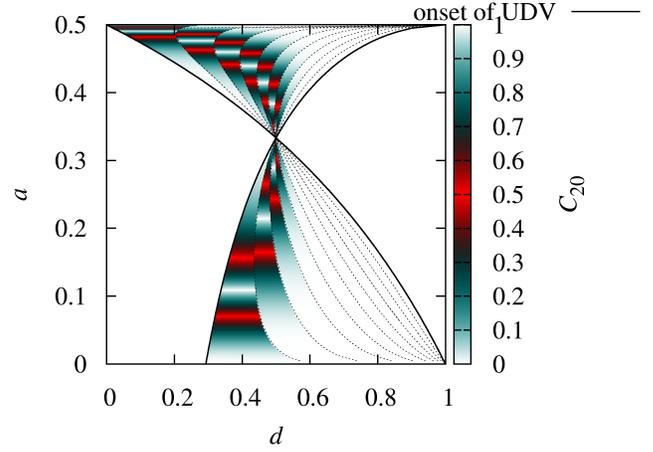}
  \caption{\label{FIG.contrast.measure} The UDV intensity, quantified by the contrast measure with $p=20$, as a function of the parameters $a$ (local dynamics) and $d$ (coupling strength). Solid lines given by eq. (\ref{EQonset.udv.curves}), indicate the transition to UDV. Dashed lines denote the transitions between the  stabilities of the fixed points da 20th iteration of the map. This figure is symmetric around $d=1$.}
\end{figure}

Since we are dealing with the magnitude of the eigenvalues, we must consider only two terms. The extremes of the spectrum of eigenvalues are then given by: $|\eta_1| = |\left(\frac{1-a}{a}\right)(1-d)|$ and $|\eta_3\eta_4| = |\frac{2-2a}{1-2a}(1-d)^2|$ \footnote{The itinerary $\overline{I_3I_4}$ represents the fixed point $x^*=1-a$ of the map (\ref{EQbungalow}). Exactly in this point the map is non-differentiable and its invariant density is discontinuous (except for $a=1/3$). However,  the density on the right and left of the point  $x^*$  are proportional to  $|\eta_3|$ and $|\eta_4|$, respectively. Thus, the itinerary  $\overline{I_3I_4}$ represents a weighted average of the dynamics, in both sides. In the remaining cases, each itinerary of size $n$ is associated with a point of period $p=n$.   }.
From (\ref{EQudvboundaries}) and solving for $d$, we have:
\begin{equation}\label{EQonset.udv.curves}
	d_c = 
	\left\{
		\begin{array}{ccc}
			1\pm\frac{a}{1-a} & \mbox{for} &  0<a<1/3\\
			1\pm\sqrt{\frac{1-2a}{2-2a}} & \mbox{for} & 1/3<a<1/2
		\end{array}
	\right.
\end{equation}
The dependece of $d_c$ on $a$ is indicated by the solid lines in Fig. \ref{FIG.contrast.measure}. Note that for  $a=1/3$ both lines intersects, indicating that there is no UDV in the system for that value of $a$. This is expected for two tent maps lineraly coupled.

The trace  $T^n$ show us the eigenvalues spectrum for a interval of time $n$, as well as the number of possible eigenvalues. We also know that the diagonal elements of $T^n$ are related to the stability coefficients (eigenvalues) of the $n$-periodic points.

Reference {\cite{PRE.v59.pR3807.y1999}} introduces the quantity
\begin{equation}\label{EQdefcontrast}
  C_p \equiv \left|\frac{\mu_2(p)-\mu_1(p)}{\mu_2(p)+\mu_1(p)}\right|, 
\end{equation}
called contrast measure, which quantifies the intensity of UDV. In Eq. (\ref{EQdefcontrast}), the quantities $\mu_i(p)$ read 
\begin{eqnarray}
  \mu_1(p) & = & \sum_k \frac{1}{|\Lambda_{\parallel}(k;p)|}\Theta(1 - |\Lambda_{\perp}(k;p)|)\label{mu1} ;\\
  \mu_2(p) & = & \sum_k \frac{1}{|\Lambda_{\parallel}(k;p)|}\Theta(|\Lambda_{\perp}(k;p)| - 1)\label{mu2} ,
\end{eqnarray}
in which $\Theta(\cdot)$ is the Heaviside function\footnote{We define here central directions ($|\Lambda_\perp| = 1$) as stable ones.}; $\Lambda_{\parallel}(k;p)$ and $\Lambda_{\perp}(k;p)$ are the eigenvalue associated with the longitudinal and transversal directions to the synchronization subspace, respectively. These eigenvalues are calculated on the $p$-periodic point labeled by $k$.  The summation extend over all fixed points of the $p$-th iteration of map. Thus, for  $p$ large enough , $\mu_{1, 2}(p)$ gives the probability of visitation of a region with unstable dimension 1 or 2 in the $p$-th iteration of the map.

Now, using what was described above, we can quantify the UDV from the coefficients in Eq.  (\ref{EQseries}) and the Eq.  (\ref{EQdefcontrast}). We must observe that the fraction of the positive tranversal Lyapunov exponents \cite{Pinto.IJBC} at $p$-finite time is {\it exactly} given by $\mu_2(p)$. This fraction is a metric dignostic for UDV. If we change $\Theta(.)$ by $\ln|\Lambda_{\perp}(k;p)|$ in (\ref{mu1}) and (\ref{mu2}), then $\mu_{1}(p)+\mu_{2}(p)$ gives $\langle \lambda_{\perp}(p) \rangle$. So, each term in summation gives the contribution for the transversal stability of $S$ of the respective UPO.

Figure \ref{FIG.contrast.measure} shows, in color scale, the intensity of UDV, quantified by the contrast measure, in the space parameter. Observing  the figure,  we notice a large region, limited by the solid lines, in which the system is non-hyperbolic ($C_p\neq 0$). There is also a large region  in which  UDV is weak. For these small values of $C_{20}$ the set of periodic orbits responsible by UDV has positive measure, but very small. Thus, a numerical diagnostic of non-hyperbolity, as the fraction of positive finite time Lyapunov exponent, typically cannot identify such regions. 


In conclusion, we have seen that the synchronized subspace of two coupled bungalow maps presents four intervals of linearity, which define four Markov partitions of the phase space. Since  UDV does not occur in the longitudinal direction of this subspace, we are able to study analitically the symbolic and interval dynamics in $S$. Pursuant to this study we found the stability coefficients of periodic points of the dynamics in the subspace synchronization, which in turn allowed us to write an exact expression for the  contrast measure.  Thus, we establish analytical solutions that show the onset of UDV, as well as the transitions between the stability of periodic points in parameter space. We can use this result to identify regions in parameter space as long as the solutions remain valid.

This work has only been able to touch on the a simple dynamical system. However, the preliminary study reported here has highlighted the need to explore the possibilities of finding analytical solutions to the problem of UDV. As this issue involves the validity of numerical solutions is important to have exact solutions for models that are studied. Clearly, further studies are needed to understand the UDV for systems with higher dimensions and arbitrary elements. To carry on this research we intend to study the UDV in a coupled map lattice whose couplings changes over time.


This work has been made possible thanks to the partial financial support from the following Brazilian research agencies: CNPq, CAPES and Funda\c c\~ao Arauc\'aria.

\bibliographystyle{apsrev4-1}
\bibliography{bungalow-v3}

\end{document}